\documentstyle[psfig,prd,aps]{revtex}
\def \pom {{I\!\!P}}

\newcommand{\rkr}{\mbox{\boldmath $r$}}

\begin{document}
\draft
\title{The QCD Pomeron in Ultraperipheral Heavy Ion Collisions: \\ II. Heavy Quark  Production}
 \author{ V.P.
Gon\c{c}alves $^{1,\dag}$\footnotetext{$^{\dag}$ E-mail:
barros@ufpel.tche.br} and  M. V. T. Machado
$^{1,2,\star}$\footnotetext{$^{\star}$ E-mail: magnus@if.ufrgs.br}
}
\address{$^1$ Instituto de F\'{\i}sica e Matem\'atica,  Universidade
Federal de Pelotas\\
Caixa Postal 354, CEP 96010-090, Pelotas, RS, BRAZIL}
\address{$^2$ High Energy Physics Phenomenology Group, IF-UFRGS\\
Caixa Postal 15051, CEP 91501-970, Porto Alegre, RS, BRAZIL}
 \maketitle

\begin{abstract}
The heavy quark production in ultraperipheral heavy ion collisions
is investigated, with  particular emphasis   on the  results from the
coherent interactions given by the  two-photon process. One
addresses the heavy quark total cross sections at photon level
considering the saturation model and the BFKL dynamics in the
color  dipole picture. The corresponding cross sections at nuclear
level are presented. It is verified that the QCD dynamics implies
an enhancement of the cross section in comparison with previous
calculations.
\end{abstract}


\section{INTRODUCTION}

       The behavior of $ep/pp$ scattering
in  the limit of high center-of-mass energy $\sqrt{s}$  and fixed
momentum transfer is one of the outstanding open questions in the
theory of the strong interactions.  In the late 1970s, Lipatov and
collaborators \cite{bfkl} established the papers which form the
core of our knowledge of Regge limit (high energy limit) of
Quantum Chromodynamics (QCD). The physical effect that they
describe is often referred to as the QCD Pomeron, with the
evolution described by the BFKL equation. The simplest process
where  this equation applies  is the high energy scattering
between two heavy quark-antiquark states, {\it i.e.} the
onium-onium  scattering. For a sufficiently heavy onium state,
high energy scattering is  a perturbative process since the onium
radius gives the essential scale at which the running coupling
$\alpha_s$ is evaluated.
 This
process was proposed as a gedanken experiment to investigate the
high energy regime of QCD in Refs. \cite{mueller,muepat,muechen}
(see also Refs. \cite{nik}). In the dipole picture \cite{mueller},
the heavy quark-antiquark pair and the soft gluons in the limit of
large number of colors $N_c$ are viewed as a collection of color
dipoles. In this case, the cross section can be understood as a
product of the number of dipoles in one onium state, the
number of dipoles in the other onium state and  the  basic cross
section for dipole-dipole scattering due to two-gluon exchange. At
leading order (LO),  the cross section grows rapidly with the
energy ($\sigma \propto \alpha_s^2  \, e^{(\alpha_{\pom} - 1)Y}$, where
$(\alpha_{\pom} - 1) = \frac{4\alpha_s\,N_c}{\pi}\,\ln 2$ and $Y =
\ln\,s/Q^2$) because the number of dipoles in the light cone
wave function grows rapidly with the energy. Therefore, in
principle, the analysis of the energy dependence of the cross
section could disentangle the QCD Pomeron effects. However, for
$ep/pp$ colliders, the study of the QCD Pomeron is made difficult
by the fact that the cross section is influenced by both short
 and long distance physics. Only when specific conditions
are satisfied is that one can expect
to determine the QCD Pomeron effects. Some examples are the
forward jet production in deeply inelastic events at low values of
the Bjorken variable $x$ in lepton-hadron scattering and jet
production at large rapidity separations in hadron-hadron
collisions.

Over the past few years much experimental effort has been devoted
towards to observe the QCD Pomeron in those processes, but the
conclusions are not unambiguous (see e.g. Ref.
\cite{reviewhighenergy}). This fact is mainly associated with the
theoretical uncertainty present in the high energy limit of QCD.
Only recently, after an effort of ten years, the next-to-leading
order (NLO) corrections to the BFKL equation were obtained (for a review on
NLO BFKL corrections, see e.g. Ref. \cite{nlosalam}  and
references therein). Moreover, since the BFKL equation predicts
that for $s \rightarrow \infty$ the corresponding cross section
rises with a  power law of the energy, violating the Froissart
bound,  new dynamical effects associated with the unitarity
corrections are expected to stop further growth of the cross
sections \cite{hdqcd}. This expectation can be easily understood:
while for large transverse momentum $k_{\perp}$, the BFKL equation
predicts that the mechanism $g \rightarrow gg $ populates the
transverse space with a large number of small size gluons per unit
of rapidity (the transverse size a gluon with momentum $k_{\perp}$ is
proportional to $1/k_{\perp}^2$), for small $k_{\perp}$ the produced gluons
overlap and fusion processes, $gg \rightarrow g$, are equally
important. Considering this process,  the rise of the gluon
distribution with transverse momenta below a typical scale, energy
dependent, called saturation scale $Q_s$ is reduced, restoring the
unitarity. It is important to salient that Golec-Biernat and
Wusthoff \cite{golec} have shown that a saturation model  is able
to describe the DESY $ep$ collider HERA data, in particular
the transition from the perturbative  to the nonperturbative
photoproduction region, including the  inclusive and diffractive
cross sections. Therefore, the description of the QCD Pomeron
still is an open question, which deserves more detailed analysis.

This situation should be improved in the future with the next
generation of linear colliders (For a recent review of BFKL
searches, see e.g. Ref. \cite{reviewhighenergy}). In particular,
future $e^+ e^-$ colliders probably will allow to discriminate
between BFKL and saturation predictions \cite{timkwiemot} (For
other analysis of photon-photon collisions see Refs.
\cite{gamagama,gambrod,gamboone}). Such reaction presents
analogies with the process of scattering of two onia discussed
above. Although the onium-onium scattering is a gedanken
experiment, off-shell
photon scattering at high energy in $e^+\,e^-$ colliders, where
the photons are produced from the leptons beams by bremsstrahlung,
plays a similar role. In these two-photon reactions, the photon virtualities can be made
large enough to ensure the applicability of the perturbative
methods. Similarly, we can expect to test the QCD Pomeron in heavy
flavor production in two-photon collisions, where the hard scale is provided
by the heavy quark mass.

In the last years many authors have studied in detail the referred
process. From the view of the color dipole picture, in Ref.
\cite{timkwiemot} the contribution of the dipole-dipole
interaction was taken into account considering a generalization
of the original saturation model \cite{golec} to the two-photon
reactions. There, the results present good agreement with data
from $e^+e^-$ colliders, remaining some room to  discuss issues
concerning resolved contributions. In Ref. \cite{szczurek}, these
questions were revisited using the saturation model, where
threshold effects were addressed in an accurate way as well as the
hadronic single-resolved contribution was computed within the
dipole formalism. A current open question is the discrepancy
between the theoretical results and the data on heavy quark
production, mostly for the bottom case, in the $\gamma p$
photoproduction and two-photon reactions. This feature is also
corroborated by the dipole-dipole approach from Ref. \cite{dondosch}.  As found
in Ref. \cite{szczurek}, even the inclusion of resolved
contributions does not remove completely the observed deficit.
Quite promising results are obtained in Ref. \cite{mottim}, where
heavy quark production in two photon reaction is calculated within
the $k_{\perp}$-factorization formalism and the  unintegrated
gluon density for the photon is determined using the  saturation
model.

>From the view of the BFKL approach, there are several calculations using the leading
logarithmic  approximation \cite{gamagama,gambrod,gamboone}  and
 considering the next-to-leading corrections to the total
cross section $\gamma^* \gamma^*$ process
\cite{gamboone,gammaNLO}. In particular, a stable next-to-leading order  program
relying on the  BLM optimal scale setting \cite{NLOBLM} has
produced  good results with OPAL and L3 data at LEP2
\cite{gammaNLO}. However, the heavy quark production from real
two-photon process is still an open question. Only the effects of charm mass in the $\gamma^* \gamma^*$ total
cross section  for the kinematical region of L3 at LEP  was
analyzed by Bartels et al. \cite{bartels} considering the LO BFKL
dynamics. We address that issue in our analysis, considering a
BFKL dipole approximation to the QCD dynamics.  Similarly to the
previous analysis on two-photon physics, one shows that  the next
generation of colliders will be able to discriminate the dynamics
associated to the QCD Pomeron.

Recently, we have proposed in Ref. \cite{vicmag}
  to investigate
QCD Pomeron effects in a different context, namely  in
photon-photon scattering at ultraperipheral heavy ion collisions.
In this case, the cross sections are enhanced since the $\gamma
\gamma$ luminosity increases as $Z^4$, where $Z$ is the atomic
number \cite{bert,reviewbaur}. There, we have analyzed the double
diffractive $J/\Psi$ production in $\gamma \gamma$ collisions,
with the photons coming from the Weizs\"acker - Williams spectrum
of the nuclei. For that process  our results have indicated that
future experimental analysis  can be useful to discriminate the
QCD dynamics at high energies. Here we extend our proposal for
heavy quark production. In principle, at   ultraperipheral heavy
ion collisions, a photon stemming from the electromagnetic field
of one of the two colliding nuclei can interact with one photon of
the other nucleus (two-photon process) or can penetrate into the
other nucleus and interact with its hadrons (photon-nucleus
process). Therefore, heavy quarks can be produced in photo-nucleus
interactions as well as photon-photon interactions. The first case
has been extensively discussed  in literature
\cite{perihe1,perihe2,perihe3} (For a review, see Ref.
\cite{perihe4}) and recently revised in different approaches
\cite{gelis,vicber,kleinvogt}. On the other hand, heavy  quark
production in two-photon collisions has been discussed essentially
in Refs. \cite{kleinvogt,soff,baur2}. In these analysis, the
underlying contribution comes from the QED quark-box diagram (QPM)
at LO [Fig. \ref{dipdiagrams} (top panel)], which dominates in the
available energies at RHIC, $W_{\gamma \gamma}\leq 6$ GeV. In Ref.
\cite{kleinvogt}, besides the direct contribution  also the
resolved contributions were carefully calculated and the latter is
shown to be negligible.  However, at the  LHC experiment, where
the two-photon energies can reach to  $W_{\gamma \gamma}\leq 200$
GeV, the QCD diagrams become important and the QCD Pomeron should
dominate in this limit  [Fig. \ref{dipdiagrams} (bottom panel)].
Consequently, an enhancement  of the cross section associated to
the dynamics is expected. We will address these issues in the
section \ref{hqperipheral}, complementing the detailed study presented in Ref.
\cite{kleinvogt}.

 Here, we
will restrict our analysis for the two-photon process and its
potentiality to investigate the QCD dynamics. Similarly as in
Ref. \cite{vicmag},  our  goal  here is twofold:  to analyze the
potentiality of this process to constraint the QCD dynamics at
high values of energy and to provide reliable estimates for the
cross sections concerning that reaction. Relativistic  heavy-ion
collisions are a potentially prolific source of $\gamma \gamma$
collisions at high energy colliders. The advantage of using heavy
ions is that the cross sections varies as $Z^4 \alpha^4$ rather
then  only   $\alpha^4$. Moreover, the maximum $\gamma \gamma$
collision energy  $W_{\gamma \gamma}$ is $2\gamma /R_A$,  about 6
GeV at RHIC and 200 GeV at LHC, where $R_A$ is the nuclear radius
and $\gamma$ is the center-of-mass system Lorentz factor of each
ion. For two-photon collisions, the cross section for the reaction
$AA \rightarrow AA Q\overline{Q}$, where $Q$ $= \,c,\,b$, will be
given by
\begin{eqnarray}
\sigma_{AA \rightarrow AA \,Q\overline{Q}}\,\,(s) = \int d\tau \,
\frac{d {\cal{L}}_{\gamma \gamma}}{d\tau} \, \hat \sigma_{\gamma
\gamma \rightarrow Q\overline{Q}}\,\,(\hat s), \label{sigfoton}
\end{eqnarray}
where $\tau = {\hat s}/s$, $\hat s = W_{\gamma \gamma}^2$ is the square of the
center of mass (c.m.s.) system energy of the two photons, $s$ of
the ion-ion system, $d{\cal{L}}_{\gamma \gamma}/d\tau$  is the
photon luminosity and $\hat \sigma_{\gamma \gamma \rightarrow
Q\overline{Q}}(\hat s)$ is the cross section of the $\gamma
\gamma$ interaction (For details related to the numerical
expressions,  see Ref. \cite{vicmag}). Our approach excludes
possible final state interactions of the produced particles with
the colliding particles, allowing reliable calculations of
ultraperipheral heavy ion collisions. Therefore, to estimate the
heavy quark  production it is only necessary to consider a
suitable QCD model for photon-photon interaction.

This paper is organized as follows. In Sec. \ref{hqphoton},  we
calculate the total cross sections for the process $\gamma \gamma
\rightarrow Q\bar{Q}$ based on the color dipole picture, using the
dipole cross section given by the saturation model and BFKL
approach. In Sec. \ref{hqperipheral}, the corresponding heavy
quark cross section in coherent ultraperipheral heavy ion
collisions are presented, with a particular emphasis in the LHC
energies. In the last section we drawn our conclusions and
summarize the main results.

\section{Heavy-quark production at the  photon level}
\label{hqphoton}

The heavy quark production via photon-photon collision is a
powerful process testing the applicability  of perturbative QCD
methods. The photon induced reactions have the advantage of being
clear concerning the probing structure and the large mass of the
heavy quarks produced allows perturbative physics to be employed.
The mechanisms of production of such process have been
systematically studied, despite the available data being modest
and having large uncertainties (see Ref. \cite{klasen} and
references therein).  An important experimental verification is
that the data on open  heavy quark pair production, mostly for
bottom quark, stay above theoretical calculations by a sizeable
factor. Such a feature has generated several studies which rely on
higher orders in the collinear factorization approach or on the
successful results based on the semihard approach.

An important approach describing two-photon process is given by
the color dipole formalism \cite{mueller,nik}. The simple physical
picture is provided by the photon splitting in a quark-antiquark
pair (virtual components in the transverse plane), called color
dipole, long after the interaction and the further  scattering of
these pairs with the target. Such an approach has produced a
consistent and unified description of high energy photon induced
processes in deep inelastic scattering  and two-photon reactions,
including the deep inelastic diffractive dissociation. In
particular, in the two-photon interactions, the target is also
given by a color dipole. The wave functions describing the photon
virtual fluctuation are completely determined from perturbative
methods, whereas the effective color dipole cross section has to
be modelled. It may contain nonperturbative and higher twist
contributions as well as parton saturation effects  as we should
discuss latter on.

\begin{figure}[t]
\centerline{\psfig{file=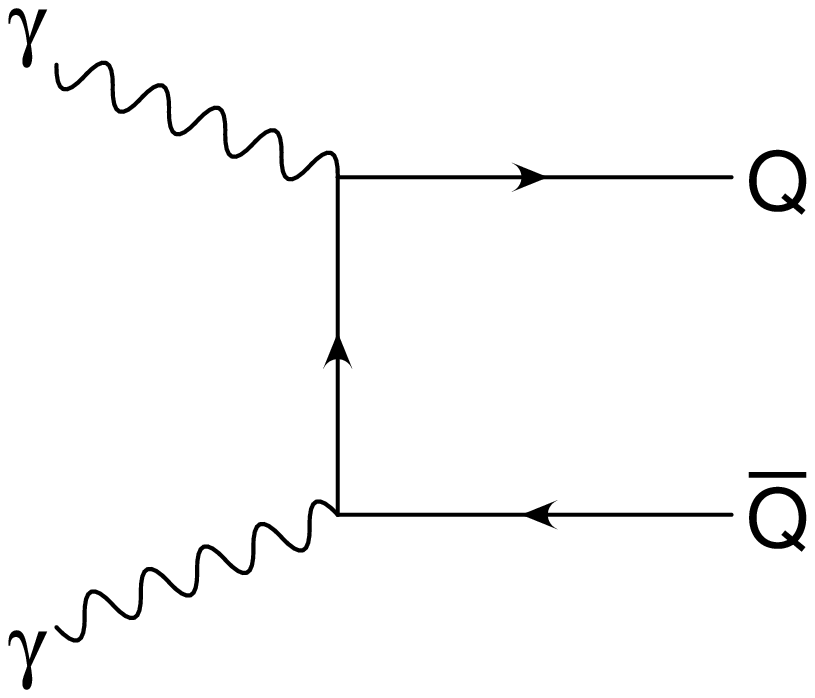,width=65mm}}
\centerline{\psfig{file=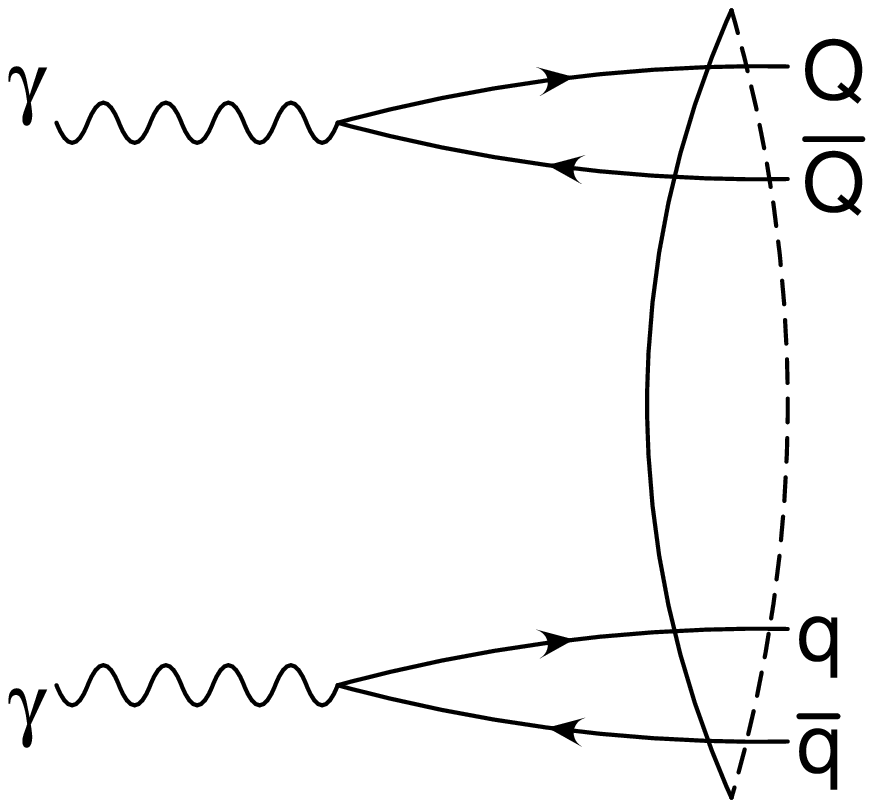,width=65mm}}
\caption{The diagrams contributing to the QPM piece (top panel)  and dipole-dipole interactions (bottom panel). Figures from Ref. [10].}
\label{dipdiagrams}
\end{figure}

The $Q\bar{Q}$ pair production mechanism can be separated into
three main contributions: (i) the quark box diagram (QPM), with
photons coupling to the same heavy quark line, which dominates at
not so large photon-photon  c.m.s.  energies, $W_{\gamma \gamma}$
[Fig. \ref{dipdiagrams} (top panel)]; (ii) the dipole-dipole
interaction, where both photons fluctuate in quark-antiquark pairs
and interact through gluonic exchange, which dominates at high
energies  [Fig. \ref{dipdiagrams} (bottom panel)]; (iii) single
and double resolved contributions, where one of the photons
fluctuates into the vector mesons ($\rho$, $\omega$, $\phi$,
$\ldots$) interacting with the remaining dipole, being a
background to the dipole-dipole reaction. The first contribution
is well known, providing a $W_{\gamma \gamma}^{-2}$ behavior
modulo logarithmic corrections, given by for example on Refs.
\cite{serbopr,donngam}. The second one will be reproduced in the
following, making use of a sound model for the dipole cross
section \cite{timkwiemot}. The third one will be disregarded, but
we quote the  study presented in Ref.  \cite{szczurek}, where
these pieces were calculated for the first time in the color
dipole approach.

The color dipole formulation gives the following expression for the heavy quark pair
production in a (real) two-photon  reactions at high energies,
\begin{eqnarray}
\sigma^{\mathrm{dd}}(\gamma \gamma \rightarrow Q\bar{Q}) & = &  \sum_{q_1\neq Q} \int |\Psi_{q_1\bar{q}_1} (\rkr_1, z_1)|^2 \,  |\Psi_{Q\bar{Q}} (\rkr_2, z_2)|^2 \, \sigma_{\mathrm{dd}}\,(\rkr_1, \rkr_2, \tilde{x}_{ab})\,  d^2 \rkr_1 \,d^2 \rkr_2 \,dz_1\, dz_2 \nonumber \\
&+ &   \sum_{q_2\neq Q} \int |\Psi_{Q\bar{Q}} (\rkr_1, z_1)|^2 \,  |\Psi_{q_2\bar{q}_2} (\rkr_2, z_2)|^2 \, \sigma_{\mathrm{dd}}\,(\rkr_1, \rkr_2, \tilde{x}_{ab})\,  d^2 \rkr_1 \,d^2 \rkr_2 \,dz_1 \,dz_2 \,,
\label{dipole_formula}
\end{eqnarray}
where $\Psi_{q\bar{q},\,(Q\bar{Q})} (\rkr, z)$ are the light (heavy) quark-antiquark
wave functions of the photon in the mixed representation. The transverse separation
(dipole size) of the respective quark pair is denoted by $\rkr$ and their longitudinal
momentum fraction denoted by $z$. The dipole-dipole cross section, $\sigma^{\mathrm{dd}}$,
depends on the dipole sizes and on the parameter $\tilde{x}_{ab}$, driving the energy behavior.

First, we write down  the expressions  for the wave functions
squared including  the respective quark helicities $T,\,L$ and
photon virtuality defined as $Q^2$,
\begin{eqnarray}
 |\Psi_T (\rkr, z)|^2 & = &  \frac{6\,\alpha_{em}}{4\,\pi^2}\,\sum_f \,e_f^2 \, [z^2 + (1-z)^2]\, \varepsilon_f^2\, K_1^2\,(\varepsilon_f\,r) + m_f^2\,  K_0^2\,(\varepsilon_f\,r)\,, \label{transwf}\\
 |\Psi_L (\rkr, z)|^2 & = &  \frac{6\,\alpha_{em}}{4\,\pi^2}\,\sum_f \,e_f^2 \, 4\,Q^2\,z^2(1-z)^2 \, K_0^2\,(\varepsilon_f\,r)\,,
\end{eqnarray}
where $e_f$ and $m_f$ stands for the charge and  mass of the quark
having flavor $f$ and $K_{0,1}$ are the McDonald-Bessel functions.
The auxiliary variable $\varepsilon$ is defined as
$\varepsilon^2=z(1-z)\,Q^2+m_f^2$ which in the real photon case,
$Q^2=0$,  considered here, simplifies to $\varepsilon=m_f$. Also
in such   case,  the longitudinal photon component is obviously
suppressed. The light quark masses are considered to be  equal and
an effective value $m_q\sim 0.2$, is  determined from fitting
two-photon data \cite{timkwiemot}.

Lets discuss now a specific model for the dipole-dipole cross
section. Here, we follows the formulation of the saturation model
applied to two-photon reactions \cite{timkwiemot}. The cross
section describing the dipole interaction satisfies the saturation
property, i.e. a constant dependence on energy at large dipole
sizes, and color transparency behavior at small dipole sizes. The
interface between this two regions is controlled  by the
saturation scale, which is energy dependent and has been
determined from fitting data  on small-$x$ deep inelastic data
\cite{golec}. Explicitly, the dipole-dipole cross  section takes
the eikonal-like form,
\begin{eqnarray}
\sigma_{\mathrm{dd}}\,(\rkr_1, \rkr_2, \tilde{x}_{ab}) = \tilde{\sigma}_0\,\left[1- \exp \left( -\frac{\bar{r}^2}{4\,R_0^2\,(\tilde{x}_{ab})} \right)  \right]\,,
\label{sigddsat}
\end{eqnarray}
with the overall normalization given by
$\tilde{\sigma}_0=(2/3)\,\sigma_0$ inspired in the quark counting
rule and where $\sigma_0$ is determined for the proton case
analysis in Ref. \cite{golec}. The effective radius $\bar{r}$  is
constructed is such way to reproduce the dipole-hadron
parameterization, i.e. $\bar{\rkr}^2\sim \rkr_1^2\, (\sim
\rkr_2^2)$ in the dipole configurations $\rkr_2^2\gg \rkr_1^2
\,(\rkr_1^2\gg \rkr_2^2)$. Here, we will consider two
possibilities giving good results as found in Ref.
\cite{timkwiemot}: (I) $\bar{r}^2= r_1^2\,r_2^2/( r_1^2 + r_2^2)$,
preferable against two-photon data; (II)
$\bar{r}^2=\mathrm{min}(r_1^2,\,r_2^2)$, preferable for charm
quark pair  production data. The fitted mass for the light quarks
takes the value $m_q=0.21$ GeV considering the effective radius
(I) and $m_q=0.23$ GeV using radius (II).  The saturation radius
$R_0$ squared  reads as,
\begin{eqnarray}
R_0^2\,(\tilde{x}_{ab}) = \left( \frac{\tilde{x}_{ab}}{x_0} \right)^{\lambda} \,\,\mathrm{GeV}^{-2}\,,
\end{eqnarray}
where the parameters $x_0$ and $\lambda$ are also determined in
the original saturation model in the proton case \cite{golec}. The
saturation scale, $Q_s(\tilde{x}_{ab})$, is obtained by the
inverse of the saturation radius described above. The effective
variable $\tilde{x}$
 depends on the two-photon energy and the quark masses and reads as,
\begin{eqnarray}
\tilde{x}_{ab} = \frac{4\,m_a^2 + 4\,m_b^2}{W^2}\,,
\label{xdef}
\end{eqnarray}
where in the heavy quark production $a$ and $b$ corresponds to a
light and a heavy (or vice versa)
 quark. For low energy (large $\tilde{x}$), the dipole cross section needs to be supplemented
 by threshold effects. Following  \cite{timkwiemot}, they are taken  into account through the
 multiplicative factor $(1-\tilde{x}_{ab})^5$, coming from dimensional-counting rules.
 However, a different procedure is adopted in Ref. \cite{szczurek}, where the threshold is considered
 by imposing the kinematical constraint $M_{q\bar{q}} + M_{Q\bar{Q}}< W_{\gamma \gamma}$ in
 the integrations at  Eq. (\ref{dipole_formula}), where $M$ is the invariant mass of the pair.
 There, also the variable $\tilde{x}$ takes a different form than Eq. (\ref{xdef}), depending
 explicitly on $z_1$ and $z_2$.

\begin{figure}[t]
\centerline{\psfig{file=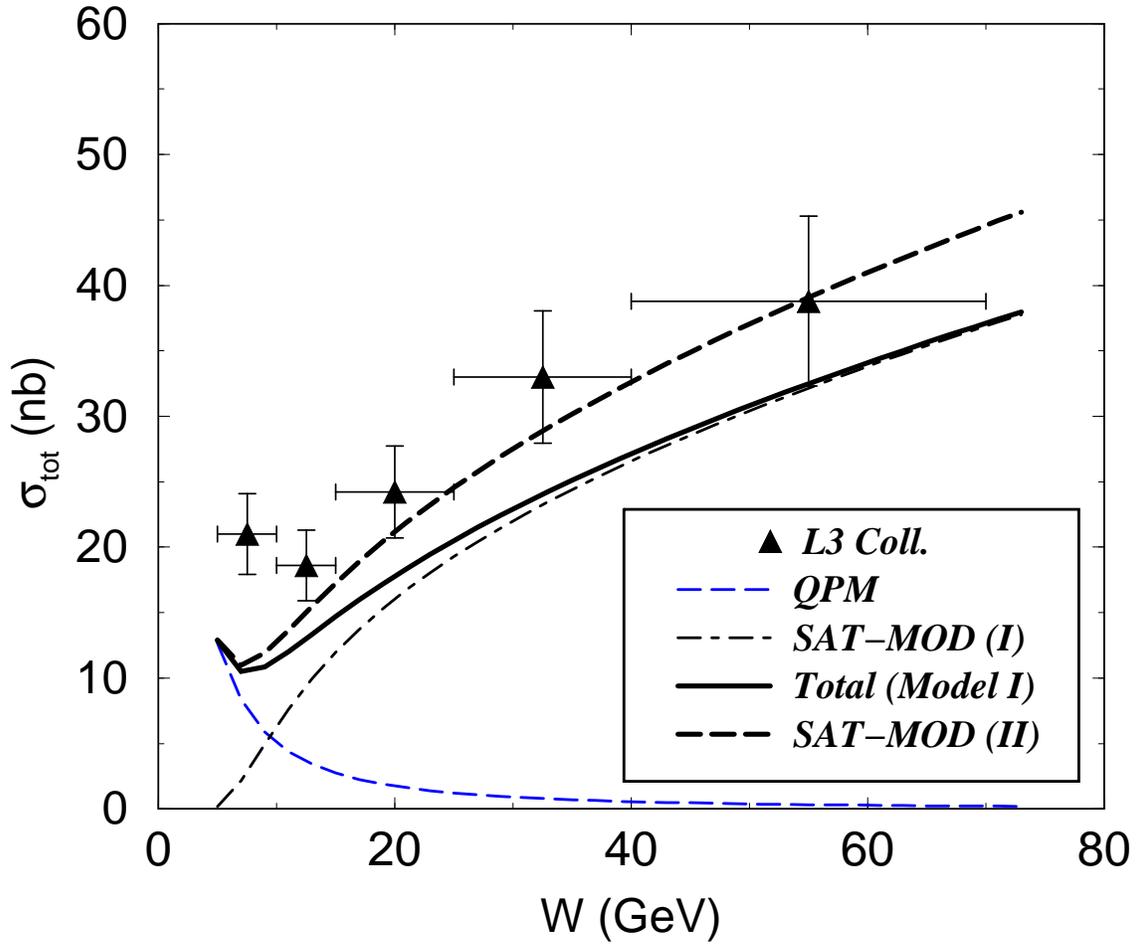,width=150mm}}
 \caption{The cross section for the inclusive charm production in two-photon reactions. The QPM contribution (thin dashed line) is shown separately. The saturation model considering different effective radius are represented by the solid (model I) and long-dashed (model II) curves. Data from L3 Collaboration.}
\label{hqpdipole}
\end{figure}

In Fig. \ref{hqpdipole}  we reproduce the results on  charm production considering the
saturation model (SAT-MOD), regarding  the quark box contribution (QPM) and the dipole-dipole
approach using the saturation model. The theoretical predictions are compared with the
experimental results from the L3 Collaboration \cite{L3}. It should be stressed that results
for the bottom case are straightforward, changing only the quark charge and mass, $e_b=1/3$
and $m_b=4.5$ GeV.  The charm mass is taken as $m_c=1.3$ GeV and the two prescriptions for the
effective radius are presented. The QPM contribution (long-dashed curve)  is very important
at low energy, starting to be negligible around $W_{\gamma \gamma}=10$ GeV. Therefore, this
contribution should dominate the ultraperipheral heavy quark production at RHIC energies, where the
maximal $W_{\gamma \gamma}$ reached  is  of order of  few GeV's. However, above
$W_{\gamma \gamma} \geq 10$ GeV, the dipole-dipole approach dominates the physical
description, providing an increase of the total cross section. The results using the
effective radius (I) (solid line) and (II) (dashed line) are presented. The contribution
from dipole-dipole interaction is shown separately for the radius (I), represented by the
dot-dashed line. This contribution plays a very important role at LHC energies, where the
 two-photon energy in ultraperipheral collisions  can reach up to $W\sim 160$ GeV.

Regarding the resolved (single and double) contributions, we have not considered  them here.
We quote Ref.  \cite{szczurek}, where these contributions are calculated using the saturation model
 within  the dipole-dipole picture. There, it was found that for the charm case the single resolved
piece constitutes about 30 \% of the main contribution. In the bottom case, these
contributions are quite smaller.

A prominent challenge in  high energy physics concerns the issue
what is the   correct dynamics driving the interaction at
asymptotic energies. It is basically dominated by gluonic
interaction and the saturation model discussed above successfully
interpolates between the soft (non-perturbative) and hard
(perturbative) regimes, whose transition relies on the parton
saturation phenomenon. However, the physics of two-photon
collisions is an optimal tool  testing the BFKL approach
\cite{bfkl}. In its dipole formulation, the BFKL dynamics is
suitable to describe the interaction of small size color dipoles
from the interacting photons. A lot of work have been done using
the LO BFKL approach and  also attempts to include NLO corrections
in two-photon reactions
\cite{gamagama,gambrod,gamboone,gammaNLO,bartels}. However, heavy
quark production in real photon-photon collisions considering the
BFKL approach was not addressed in the current literature.  Below,
we present our results for this case.

To estimate the heavy quark production in the BFKL approach we
need a  model for the dipole-dipole cross section. In particular,
the BFKL dipole cross section can be obtained from the momentum
representation and it can  be considered as a solution of the
linearized Balitsky-Kovchegov equation in the coordinate space
\cite{hdqcd,itakura}.
 The LO BFKL solution is completely known
in the momentum space, providing a steep energy dependence associated with the hard Pomeron
intercept $1+\omega_{\pom}$ and the typical diffusion feature.  For our purpose
here, we will consider the following expression for the dipole-dipole cross section from a
BFKL dynamics,
\begin{eqnarray}
\sigma_{\mathrm{dd}}\,(\rkr_1, \rkr_2, \tilde{x}_{ab}) = \frac{4\,\pi^2}{3}\,{\cal{C}}\,\left( \frac{\bar{r}^2}{\bar{r}_0^2} \right)^{\frac{1}{2}} \, \frac{e^{\,\omega\,\bar{\alpha}_s \ln (1/\tilde{x}_{ab})}}{\sqrt{2\,\pi\,\tilde{\omega}\, \bar{\alpha}_s\,\ln (1/\tilde{x}_{ab})}} \, \exp \left[\frac{-\ln^2\,(\bar{r}^2/\bar{r}_0^2)}{2\,\tilde{\omega}\,\bar{\alpha}_s\,\ln (1/\tilde{x}_{ab})}  \right]\,(1- \tilde{x}_{ab})^5 \,,
\label{sigddbfkl}
\end{eqnarray}
where $\bar{\alpha}_s=N_c\,\alpha_s/\pi$, with number of color
$N_c=3$. Moreover, for light quarks, ${\cal{C}}=1.0$,  $\bar{r}_0^2=1/\Lambda_{\mathrm{QCD}}^2$, where
one has used $\Lambda_{QCD}=0.23$. Still, $\omega\equiv 2\psi (1)
- 2\psi(1/2)=4\,\ln\,2$ and $\tilde{\omega}=28\, \xi (3)$. The
Pomeron intercept is given by
$\omega_{\pom}=1+\bar{\alpha}_s\,\omega$. The  prescription used
here is not unique. For instance, for dipole-dipole  collisions,
 the choice would be    $\bar{r} = \mathrm{min} (\rkr_1,\rkr_2)$  and
 $\bar{r}_0 = \mathrm{max} (\rkr_1,\rkr_2)$, having  impact on the final results.

\begin{figure}[t]
\centerline{\psfig{file=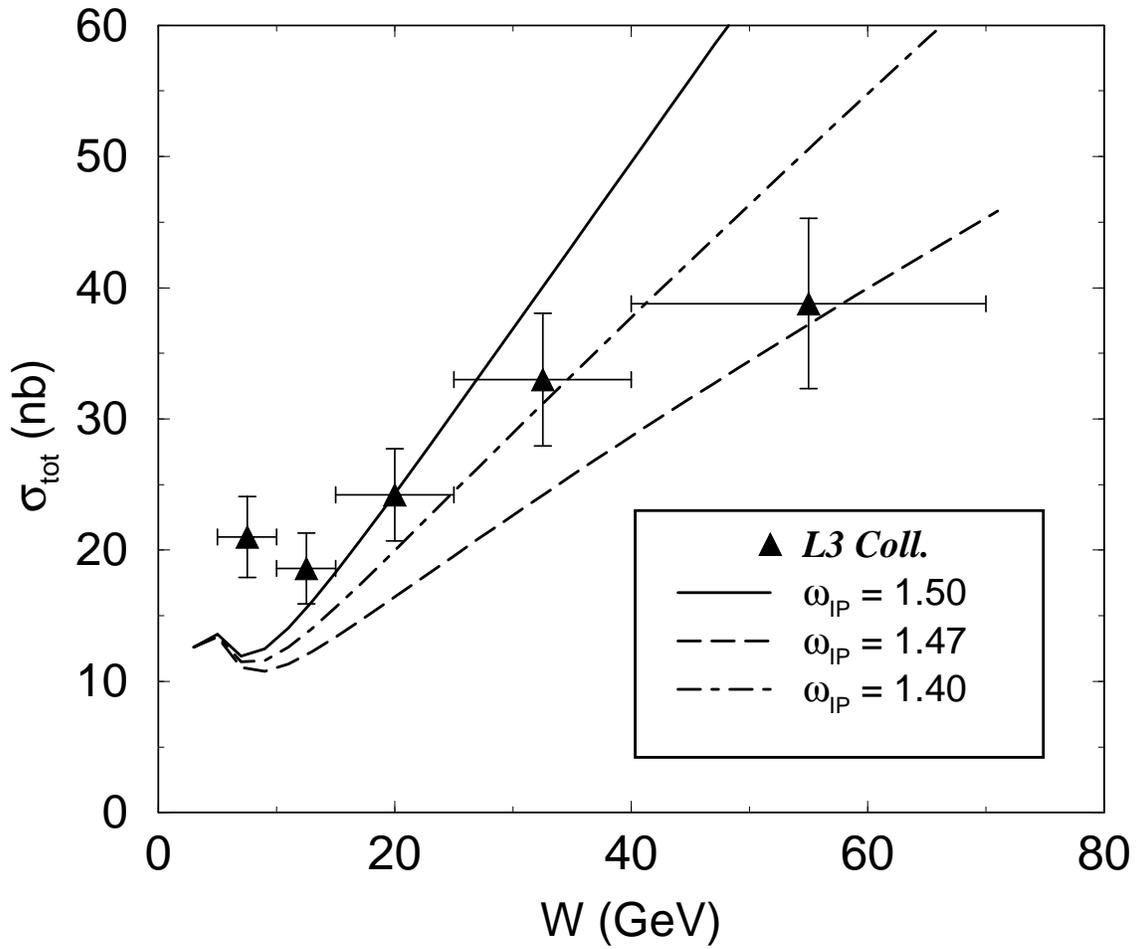,width=150mm}}
 \caption{The cross section for the inclusive charm production in two-photon process considering the BFKL dipole-dipole approach. The results for three different values for the Pomeron intercept $\omega_{\pom}$ are presented. The QPM contribution is also included. Data from L3 Collaboration.}
 \label{hqpbfkl}
\end{figure}

In  Fig. \ref{hqpbfkl} are presented the results using the
dipole-dipole cross section of Eq. (\ref{sigddbfkl}), considering
the effective radius (I) and  taking into account three different
values for the Pomeron intercept. The QPM contribution was also
included. In our  further analysis one considers the intermediate
value $\omega_{\pom}=1.47$, which is steeper than the results from
the saturation model. It should be stressed that further
phenomenology can be done, for instance taking the overall
normalization, the scale $\bar{r}_0$  and the Pomeron intercept
$\omega_{\pom}$ as free parameters. The differences  between the
predictions of the saturation model and the BFKL approach are
already sizeable in the energy interval presented here, with
important implications in the future TESLA experiment.

\begin{figure}[t]
\centerline{\psfig{file=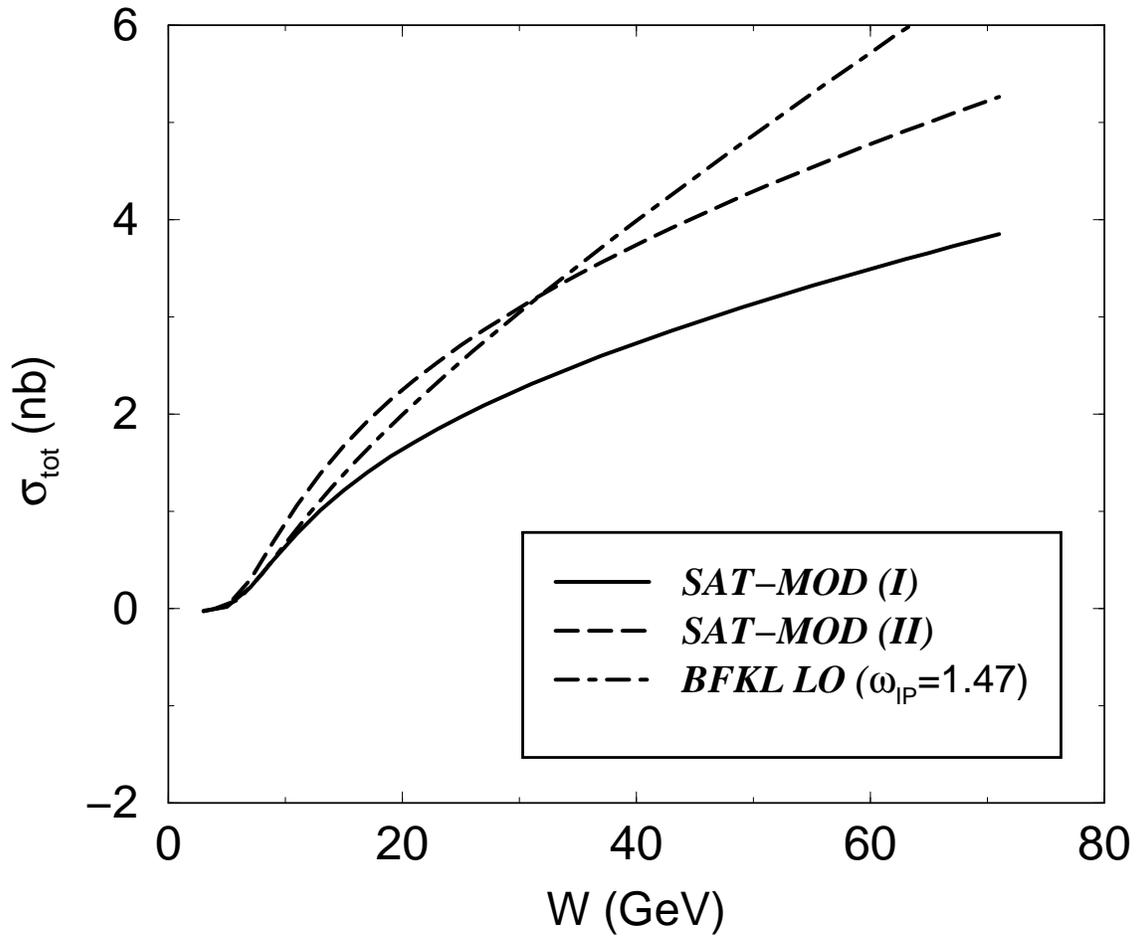,width=150mm}}
\caption{The cross section $\sigma(\gamma \gamma \rightarrow 2c\,2\bar{c})$  considering the saturation model (models I and II) and the BFKL approach, with a Pomeron intercept $\omega_{\pom}=1.47$. }
\label{doubleprod}
\end{figure}

As a final analysis at the photon level, we  should calculate the
cross section for the interesting final state $2Q\,2\bar{Q}$,
which is by definition an onium-onium scattering.  In this case
the light quarks in the bottom panel of the Fig.
\ref{dipdiagrams} are replaced by a pair of heavy quarks. The
production of this particular configuration was for the first time
addressed by Szczurek in Ref. \cite{szczurek}, calling the
attention that the dipole-dipole approach gives a unique
prediction to production of two identical heavy quarks and two
identical heavy antiquarks in the final state.  The process
$\gamma \gamma \rightarrow 2Q\,2\bar{Q}$ can only to be produced
in next-to-leading order calculation in the collinear approach
and/or hadronization process.  The cross section for this process
is given by,
\begin{eqnarray}
\sigma^{\mathrm{dd}}(\gamma \gamma \rightarrow 2Q\,2\bar{Q})  =  \int |\Psi_{Q\bar{Q}} (\rkr_1, z_1)|^2 \,  |\Psi_{Q\bar{Q}} (\rkr_2, z_2)|^2 \, \sigma_{\mathrm{dd}}\,(\rkr_1, \rkr_2, \tilde{x}_{ab})\,  d^2 \rkr_1 \,d^2 \rkr_2 \,dz_1\, dz_2 \,.
\label{doubleqq}
\end{eqnarray}
 In Fig. \ref{doubleprod} are shown the results using the Eq. (\ref{doubleqq}) and
considering the saturation model (radius I and II)  and the BFKL
dipole-dipole cross section for the charm quark case. In this case, $\bar{r}_0^2=1/m_c^2$ and we assume ${\cal{C}}= 1/9$ to reproduce the results in the light quark case.  Once again,
the BFKL result is steeper on energy than the dipole model
predictions. Similarly to Ref. \cite{szczurek}, we obtain that at high energies the cross section for $2c2\overline{c}$ is about  9 $\%$ of that for the single $c\overline{c}$ production. It is important to salient that in the calculations of the inclusive $c\overline{c}$ production cross section, the contribution of this process should be doubled because each of the quarks can  potentially be identified experimentally (See discussion in Ref. \cite{szczurek}). In next section, we present our results for heavy quark production in ultraperipheral heavy ion collisions, considering both contributions for the inclusive heavy quark production cross section.

\section{The heavy quark production in ultraperipheral collisions}
\label{hqperipheral}

Having determined the cross sections for heavy quark  production
at the photon level, in the following the results for
ultraperipheral heavy ions collisions will be presented. We give
emphasis on the LHC region, where $W_{\gamma \gamma}\leq 200$ GeV,
and a luminosity ${\cal L}=4.2\times 10^{26}$ cm$^{-2}$s$^{-1}$
for PbPb collisions ($\sqrt{s}=5500$ GeV). Details on the
effective two-photon luminosity used here can be found in the
previous work in Ref. \cite{vicmag}. As referred above, one has
been taken $m_c=1.3$ GeV and $m_b=4.5$ GeV. Further, we discuss
the mass dependence of our results.

Our predictions for the inclusive heavy quark cross section and double $c\overline{c}$ pair production at
$\sqrt{s}=5500$ GeV are presented in Table \ref{tab1}. In
comparison with the predictions from Ref.  \cite{kleinvogt}, our
results for the inclusive cross section  are higher by approximately one factor of 1.8. This
enhancement is directly associated to the QCD dynamics, which
resums higher order diagrams beyond those considered previously
in the literature. It is important to salient that this factor is
larger than the uncertainty associated with the treatment of the
$\gamma \gamma$ luminosity, which is estimated to be about 15$\%$
\cite{kleinvogt}. For RHIC energies (not presented in the Table),
due to the limited center of mass energy of the photon-photon
process, the QCD Pomeron effects are not important, and our
results are similar with those presented in  Ref.
\cite{kleinvogt}.

\begin{figure}[t]
\centerline{\psfig{file=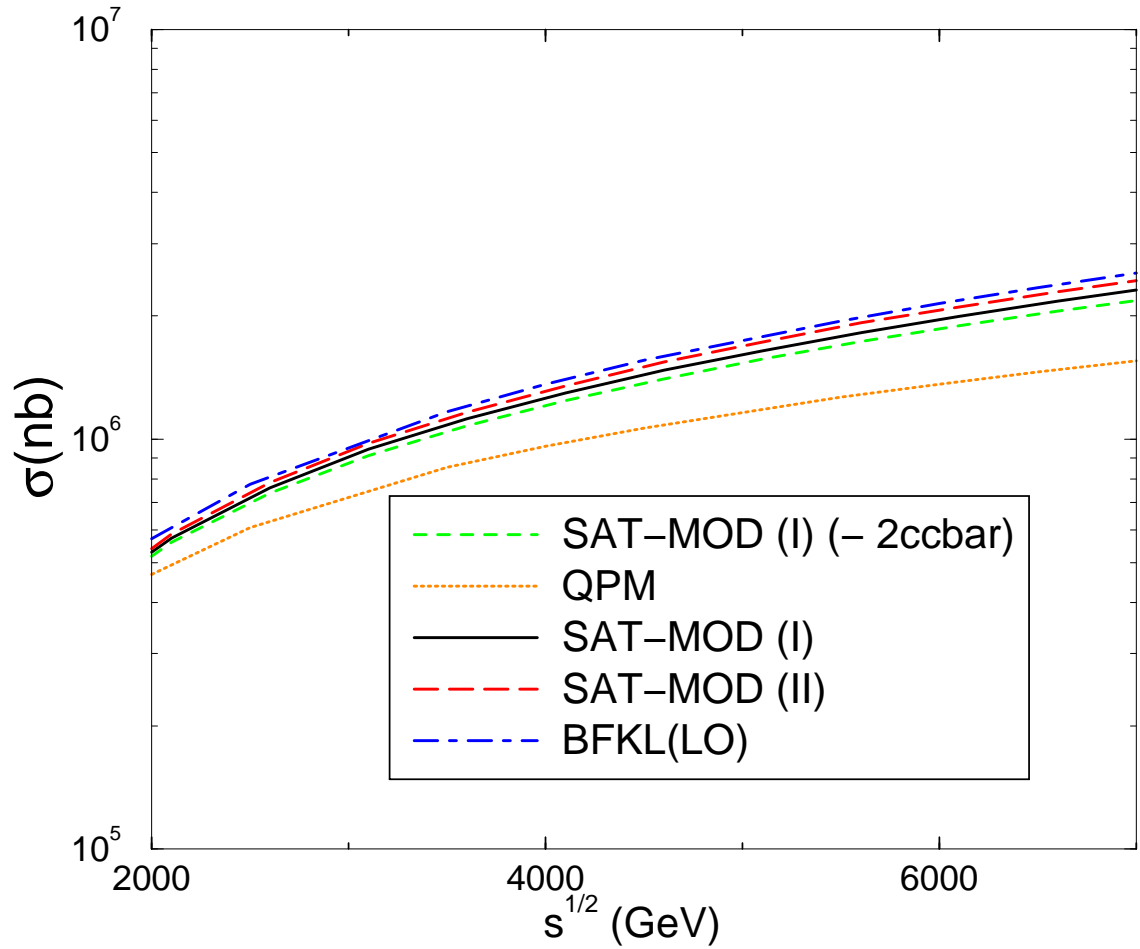,width=150mm}} \caption{Inclusive charm
 pair production in  ultraperipheral heavy ion collisions considering the distinct QCD approaches for the $\gamma \gamma$ interaction. For comparison the QPM (dotted line) and single $c\overline{c}$ (dashed line) cross section  are also presented. In the last case, the contribution of double $c\overline{c}$ production for the inclusive cross section  was subtracted.}
\label{ccbarheavy}
\end{figure}

In Fig. \ref{ccbarheavy} are presented the results for  the energy
dependence of the inclusive charm pair production in two-photon
ultraperipheral heavy ion collisions. The estimates using the
saturation model are shown as the solid and long-dashed lines,
corresponding to the choice (I) and (II) for the effective radius,
respectively. The dipole BFKL calculation is shown in the dot-dashed curve, where we have considered the Pomeron intercept
$\omega_{\pom}=1.47$. For comparison, the QPM  and the single $c\overline{c}$ cross sections are also presented. Despite the deviations between the
approaches being large at the photon level, the results for the
peripheral reaction are quite similar. This feature is expected,
since the two-photon luminosity in nuclear collisions  strongly
suppress high  energy contributions in comparison with the
behavior present in  $e^ +e^-$ collisions. Concerning the quark
mass dependence of our results, we  have checked that if one uses
$m_c = 1.2 \,$ GeV, as done in \cite{kleinvogt},  the cross
section grows approximately 25 $\%$. Therefore, for this quark
mass choice,  we predict an enhancement of a factor two. Here we
have used  $m_c = 1.3$ GeV, since with this choice the
experimental data are well described in the dipole approach.

\begin{figure}[t]
\centerline{\psfig{file=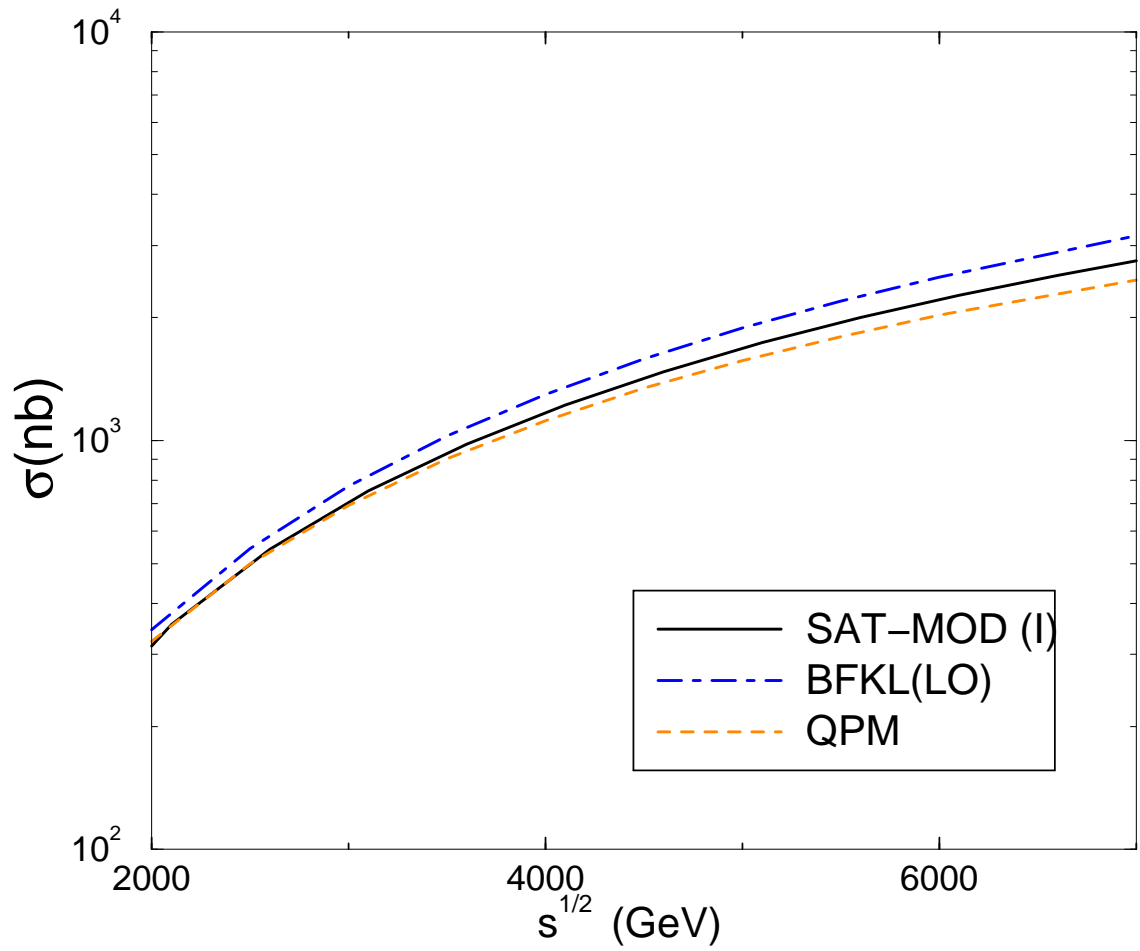,width=150mm}} \caption{Inclusive bottom
 pair production in ultraperipheral heavy ion collisions. For comparison the QPM prediction is also presented.}
\label{bbbarheavy}
\end{figure}

In Fig. \ref{bbbarheavy} are shown the results for the energy
dependence of the  inclusive bottom pair production in two-photon
ultraperipheral heavy ion collisions. The solid line represents
the estimate obtained with the saturation approach (model I), 
the dot-dashed line the dipole BFKL result and the dashed line is the QPM prediction. The deviation between
the saturation and BFKL approaches reaches up to 10\%. Similarly to the charm, the
analysis of the bottom quark production in ultraperipheral heavy
ion collisions would not allow to discriminate between the
distinct kinds of dynamics. However, these effects cannot be
disregarded at LHC energies, since our predictions are a factor
two larger than those obtained in \cite{kleinvogt}.

A comment is in order here. In  \cite{kleinvogt} a detailed
analysis of the experimental separation between photoproduction
and  two-photon interactions was presented. There, the authors
have estimated that the two-photon cross sections are at least
1000 smaller than the photoproduction cross section, which makes
the experimental separation between the two interactions very
hard. Our calculations  indicate that the inclusion of the QCD
Pomeron effects implies higher cross sections at two-photon level
and, consequently, larger cross sections in ultraperipheral
collisions. Therefore, the inclusion of these effects  implies that
the contribution of heavy quark production in two-photon
interactions is  non-negligible. However, the experimental separation
of two-photon process is still remains a challenge.

\begin{figure}[t]
\centerline{\psfig{file=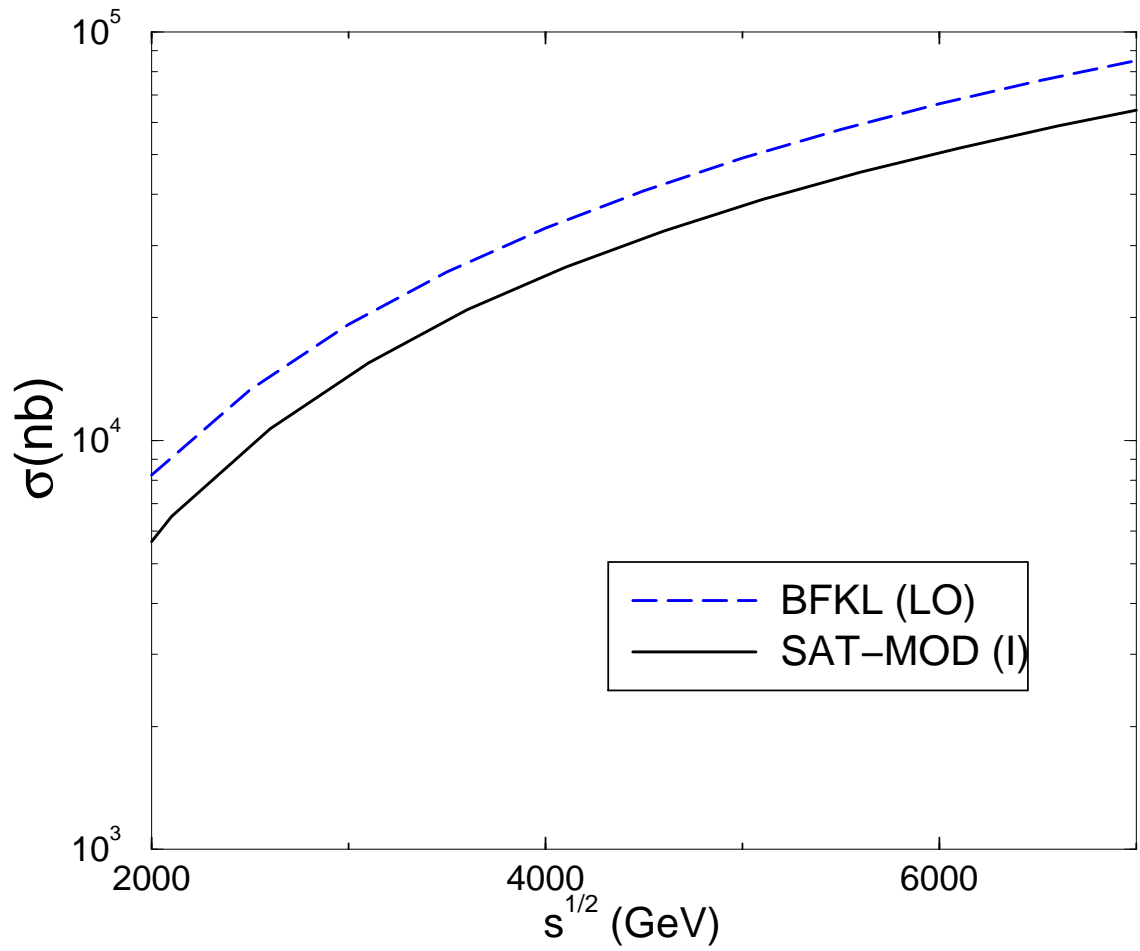,width=150mm}} \caption{Double
 pair charm production in (two-photon) Ultraperipheral Heavy Ion Collisions.}
\label{dccbarheavy}
\end{figure}

In Fig. \ref{dccbarheavy} we present  separately the results for the  double
charm pair  production in two-photon ultraperipheral heavy ion
collisions, obtained from Eq. (\ref{doubleqq}). The solid line
represents the estimate obtained with the saturation approach
(model I) and the dot-dashed line is the dipole BFKL result. The
difference  between the predictions of the  two approaches is about $20 \%$ in this
particular case. In comparison with inclusive charm pair production, the
cross section is two order of magnitude smaller, but still
sizeable for an experimental analysis. We salient  that  the efficiency of  flavor tagging is in general very small, which also makes the experimental 
detection of this process a challenge. However, it important to emphasize
that distinctly from the single pair production which is one
background of the photoproduction process, the double heavy quark
production can be produced in photo-nucleus reactions only if the
multiple interactions are considered. Since, in general, these
interactions are suppressed by a factor of about  1000
\cite{kleinnystra}, we have that for this process two photon
interactions and photon-nucleus interactions would be of same
order.

\section{Summary}
\label{conc}

We have investigated the heavy quark production in the ultraperipheral
heavy ion collisions from  two-photon reactions. The total cross sections at
the photon level are calculated using the  saturation model, describing the dipole-dipole
interaction at high energies including in a suitable way the saturation phenomenon.
The dipole BFKL approach is also considered as an important test of QCD dynamics and
predictions for the photon-photon cross section using  this approach are presented.
Both calculations are in reasonable agreement with the current experimental measurements.
In particular, we have considered a typical final state allowed by the color dipole picture,
namely the double heavy quark pair  production $\gamma \gamma \rightarrow 2Q\,2\bar{Q}$.

The resulting cross sections in ultraperipheral heavy ion
collisions are estimated  in these distinct dynamical approaches
for the QCD Pomeron and presented. Our results demonstrate that
the analysis of the heavy quark production in ultraperipheral
heavy ion collisions would not allow to discriminate between the
distinct dynamics. However, the QCD Pomeron effects cannot be
disregarded at LHC energies, since the  cross sections are
enhanced by a factor two  for these effects. Moreover, we have
analyzed the double charm pair production and verified that the
cross section for this process is sizeable, mainly due to the fact
that photoproduction background can only be  produced by
higher-order multiple interactions. Our results motivate more
detailed studies. In particular, the analysis of the mass and
transverse momentum distributions could be useful to discriminate
between photon-Pomeron  and two-photon processes. Work in this
direction is in progress.

\section*{ACKNOWLEDGMENTS}
The authors thank S. Klein for valuable  suggestions and comments. We are particularly grateful to L. Motyka for his careful reading of the manuscript and for his helpful comments. M.V.T.M. thanks the support of the  High Energy Physics Phenomenology Group (GFPAE, IF-UFRGS) at the Institute of Physics, Porto Alegre. This work was partially financed by the Brazilian funding agencies CNPq and
FAPERGS.

\newpage

\begin{table}[t]
\begin{center}
\begin{tabular} {|l|l|l|l|}
\hline
   & SAT-MOD (I) & SAT-MOD (II) & LO BFKL \\
\hline
 $c\bar{c}$ &  1810000 nb & 1900000 nb & 1951200 nb \\
\hline
 $b\bar{b}$ & 2000 nb   & - & 2200 nb  \\
\hline
 $2c\,2\bar{c}$ & 40000 nb & -   & 48000 nb  \\
\hline
\end{tabular}
\end{center}
\caption{The heavy quark production total cross sections for
ultraperipheral heavy ion collisions at LHC ($\sqrt{s}=5500$ GeV)
for PbPb.} \label{tab1}
\end{table}

\newpage

\newpage

\newpage

\newpage

\newpage

\newpage

\newpage

\end{document}